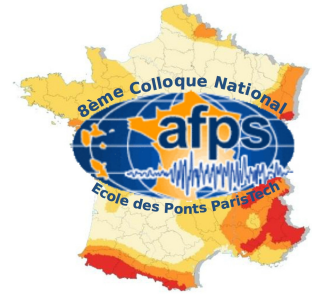

# Deux approches différentes dans l'évaluation analytique du potentiel endommageant des tremblements de terre: application et validation pour les forts séismes de Vrancea, Roumanie

**Iolanda-Gabriela Craifaleanu\*,\*\* — Ioan Sorin Borcia\*\***

*\* Université Technique de Constructions de Bucarest,*

*Département de constructions en béton armé*

*122-124 Blvd. Lacul Tei, RO-020396 Bucarest, Roumanie*

*i.craifaleanu@gmail.com*

*\*\* Institut National de Recherche et Développement dans la Construction, l'Urbanisme et le*

*Développement Territorial Soutenable – « URBAN – INCERC », Branche INCERC Bucarest, Roumanie*

*266 Şos. Pantelimon, RO-021652 Bucarest, Roumanie*

*isborcia@yahoo.com*

RÉSUMÉ. *L'article présente une comparaison entre les résultats des évaluations analytiques du potentiel endommageant des mouvements sismiques, effectuées à la base de deux paramètres différents : l'indicateur de dommages de Park et Ang et l'intensité sismique instrumentale de Horea Sandi. L'indicateur de dommage de Park et Ang est largement utilisé pour évaluer les dommages structurels grâce à sa simplicité et à sa calibration par rapport aux données expérimentales. L'intensité instrumentale a l'avantage d'être calibrée par rapport à l'échelle d'intensité EMS-98. Sur la base du traitement des accélérogrammes enregistrées pendant les forts tremblements de terre générés par la source sismique Vrancea en 1977 (magnitude de moment $M_w$ = 7,4) et 1986 ($M_w$ = 7,1), on détermine les valeurs des paramètres considérés et on construit des cartes de leur distribution spatiale. Les résultats des deux approches sont comparés entre eux, ainsi qu'avec les données fournis par les rapports concernant les dégâts constatés effectivement au bâtiment, dû aux séismes considérés.*

ABSTRACT. *The paper presents a comparison between the results of the analytical assessments of the damage potential of seismic ground motions, based on two different parameters: the Park-Ang damage index and the Sandi instrumental seismic intensity. The Park-Ang damage index is largely used for structural damage assessment due to its simplicity and its calibration against experimental data. The instrumental intensity has the advantage of being calibrated with respect to the EMS-98 intensity scale. Based on the processing of ground motions recorded during the strong earthquakes generated by the Vrancea seismic source in 1977 (moment magnitude $M_w$=7,4) and 1986 ($M_w$=7,1), the values of the considered parameters are determined and maps of their spatial distribution are generated. The results of the two approaches are compared to one another, as well as with the information provided by the reports concerning the effective damage observed in buildings due to the considered earthquakes.*

MOTS-CLÉS : *indicateur d'endommagements, intensité sismique, échelle d'intensité, EMS-98, séismes de Vrancea, cartes d'endommagement.*

## 1. Introduction

La séismicité de Roumanie est générée principalement par la source sismique de Vrancea. Cette source a généré, au fil des années, un nombre important d'événements sismiques destructifs. Selon le catalogue "Romplus" de l'Institut National Roumain de la Physique du Globe (INFP, 2009), pendant le XX$^{ème}$ siècle en Vrancea se sont produit 5 séismes avec la magnitude de moment $M_w \geq 7$, à savoir en 1901, 1908, 1940, 1977 et 1986.

Le séisme le plus destructif des temps modernes s'est produit le 4 mars 1977 ($M_w$=7,4, profondeur hypocentrale $h$ = 94 km), causant 1578 morts et environ 11300 blessés. Environ 35000 familles ont resté sans abri et les dégâts totaux ont été estimés à plus de 2 milliards de dollars. A Bucarest, 33 bâtiments multi-étagés se sont effondrés et 1424 morts ont été rapportées (Balan et al., 1982, Berg et al., 1980).

Neuf ans plus tard, le séisme de 30 août 1986 ($M_w$ = 7,1, $h$ = 133 km), a provoqué des dommages faibles ou modérés en Roumanie, en causant, pourtant, 2 morts et 558 blessés. Dans la République de Moldavie, les effets ont été beaucoup plus importants ; les dégâts ont été chiffrés à environ 680 millions de dollars (Zaicenco et al., 2004). Quatre bâtiments se sont effondrés dans la capitale, Chisinau ; des dommages ont été également enregistrés à Cahul et dans d'autres villes. Le séisme a causé deux morts et 561 blessés ; 14000 personnes sont restées sans abri.

L'article présente une évaluation analytique du potentiel endommageant des deux séismes mentionnés, en utilisant, en parallèle, deux approches différentes. La première approche est basée sur l'indicateur de dommages Park-Ang, tandis que la seconde utilise l'intensité sismique instrumentale proposée par Horea Sandi. Les résultats obtenus sont comparées entre eux, ainsi qu'avec les informations concernant l'endommagement enregistré effectivement aux bâtiments pendant les deux séismes.

## 2. Evaluation basée sur l'indicateur d'endommagement Park-Ang

### 2.1. *Spectres d'endommagement*

L'indicateur d'endommagement Park-Ang (Park et al., 1985), *DM*, est utilisé largement dû à sa calibration expérimentale et à la simplicité de son expression analytique. L'indicateur est défini par la relation suivante:

$$DM = (u_{\max}/u_{mon}) + \beta \left(E_H / F_y u_{mon}\right) \quad \text{ou} \quad DM \cdot \mu_{mon} = \mu_{max} + \beta \left(E_H / F_y u_y\right) \quad [1]$$

où $u_{max}$ est le déplacement horizontal maximal pendant le mouvement sismique, $u_{mon}$ est le déplacement capable du système sous des forces horizontales qui croissent de façon monotone, $E_H$ est l'énergie hystérétique, $F_y$ est la force de plastification, $u_y$ est le déplacement de plastification, $\mu_{max}$ est la ductilité maximale atteinte pendant le mouvement sismique, $\mu_{max} = u_{max}/u_y$, $\mu_{mon}$ est la ductilité maximale sous charges monotones, $\mu_{mon} = u_{mon}/u_y$, et $\beta$ est une constante dépendante des caractéristiques structurales, qui peut être considérée égale à 0,15 (Cosenza et al., 1993).

Pour décrire l'endommagement de la structure, on utilise l'interprétation suivante des valeurs de *DM* (Park et al., 1987, Teran-Gilmore, 1996) : *DM* = 0,4 (0,5) est considérée comme la limite supérieure de l'endommagement réparable, les valeurs situées entre 0,4 (0,5) et 1 caractérisent l'endommagement sévère, irréparable, tandis que les valeurs supérieures à 1 correspondent à la ruine du bâtiment. La valeur 0,2 este considérée comme la limite supérieure de l'endommagement insignifiant.

En vue de l'étude, on a calculé les spectres de l'indicateur d'endommagement Park-Ang, *DM*, pour les composantes horizontales des accélérogrammes enregistrés en champ libre aux séismes de 1977 et 1986. Il est à mentionner que, pour le séisme de 1977, le plus fort événement sismique de Vrancea enregistré avec des instruments modernes, on dispose d'un seul accélérogramme tridimensionnel. Cet accélérogramme a été enregistré à la station sismique INCERC Bucarest.

Comme paramètre des courbes spectrales on a choisi le coefficient de résistance $C_y$, donné par la formule

$$C_y = F_y/G \qquad [2]$$

où *G* est le poids du système. Les spectres ont été déterminés en considérant un modèle bilinéaire, élastique - parfaitement plastique, et une fraction d'amortissement critique de 5%. Le coefficient $C_y$ représente une mesure simple de la capacité de résistance du système aux forces horizontales. Il peut être mis facilement en relation avec le coefficient sismique, $C_s$, défini comme le rapport entre la force sismique spécifiée par la norme et le poids du bâtiment. En notant le coefficient de sur-résistance avec $R_{OVS}$, on peut écrire:

$$C_y = C_s R_{OVS} \qquad [3]$$

La fig. 1 présente des spectres de l'indicateur Park-Ang pour la composante nord-sud de l'accélérogramme enregistré à INCERC Bucarest le 4 mars 1977. Les spectres sont calculés en considérant une ductilité $\mu_{mon} = 8$. On remarque les valeurs *DM* très élevées qui résultent à périodes courtes et pour valeurs basses du coefficient de résistance $C_y$. Ce trait est une caractéristique générale, spécifique aux spectres du *DM*, et qui est due à sa dépendance de la ductilité. Dans la figure mentionnée on a représenté aussi les valeurs caractéristiques de l'indicateur *DM* : la limite du comportement élastique, ici à $1/\mu_{mon} = 0,125$, et les limites de l'endommagement insignifiant (*DM* = 0,2), réparable (*DM* = 0,4) et de la ruine (*DM* = 1).

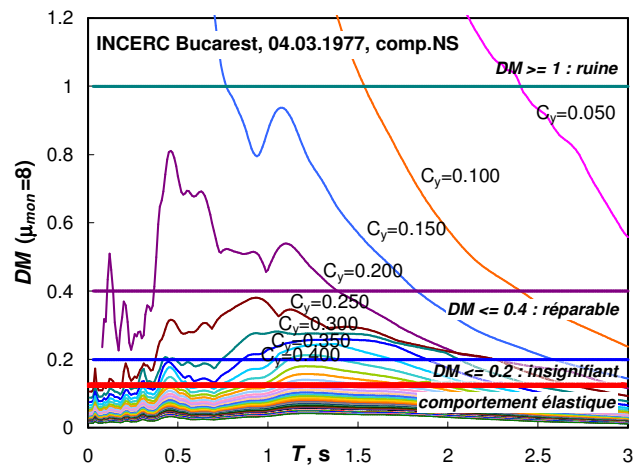

**Figure 1.** *Spectres de l'indicateur Park-Ang pour la composante nord-sud de l'accélérogramme enregistré à INCERC Bucarest le 4 mars 1977* ($\mu_{mon} = 8$)

### 2.2. *Cartes d'endommagement*

Les valeurs spectrales de *DM* ont été cartographiées, pour le séisme de 30 août 1986. Avec une magnitude de moment $M_w = 7,1$, ce séisme est le plus fort événement de la source Vrancea pour lequel on dispose d'enregistrements dans plusieurs stations, grâce à l'extension des réseaux sismiques en Roumanie après le séisme de 1977. Ces stations, situées en champ libre, sont, pour la plupart, distribuées dans la zone extra-carpatique.

Pour donner aux estimations plus de généralité, ainsi que pour des raisons tenant à la comparaison avec les résultats de la deuxième approche utilisé dans l'étude, on a calculé la moyenne des valeurs *DM* sur 3 intervalles de période : $T = 0,5…0,7$ s, $0,7…1,0$ s et $1,0…1,4$ s. Ces valeurs correspondent aux bâtiments de hauteur moyenne et élevée, qui seront analysées dans la partie finale de cet article.

En partant des valeurs cartographiées, on a généré des surfaces d'interpolation et des contours de valeur $C_y$ constante, pour avoir une vue synthétique de la distribution spatiale du potentiel endommageant.

Les figures 2 et 3 présentent les cartes des valeurs spectrales de *DM*, calculées respectivement pour $C_y = 0,10$ et $C_y = 0,20$.

En analysant les cartes, on peut remarquer que les plus grandes valeurs *DM* apparaissent sur une ligne qui unit les stations Valenii de Munte (VLM1) et Chisinau (CHS1), et qui passe aussi par l'épicentre du séisme. Des valeurs élevées sont aussi observables aux stations Focsani (FOC1) et Barlad (BIR1). Les valeurs diminuent assez rapidement avec la distance par rapport à cette ligne, de telle façon que, sur toutes les cartes, aux stations situées à son nord-ouest et à son sud–est, *DM* est bien au-dessous de la limite d'endommagement insignifiant. La région de la capitale, Bucarest, située à l'extrémité sud-ouest de la zone analysée, est caractérisée par des valeurs *DM* intermédiaires, dans tous les cas.

Une autre caractéristique commune à tous les cartes est la diminution de *DM* avec la période *T*. Cette caractéristique est également observable dans la fig. 1.

En ce qui concerne la variation avec le coefficient de résistance $C_y$, on observe une diminution substantielle des valeurs de *DM* avec l'augmentation de $C_y$ de 0,10 (fig. 2) à 0,20 (fig. 3). Etant donné la signification des grandeurs analysées, cette tendance, elle aussi observable dans la fig. 1, exprime la réduction du degré d'endommagement avec l'augmentation de la capacité de résistance.

Du point de vue qualitatif, la distribution spatiale des valeurs du *DM* reflète de manière satisfaisante la distribution réelle des zones où se sont produit des dégâts aux bâtiments ; du point de vue quantitatif, les valeurs dépassent, en général, les effets constatés.

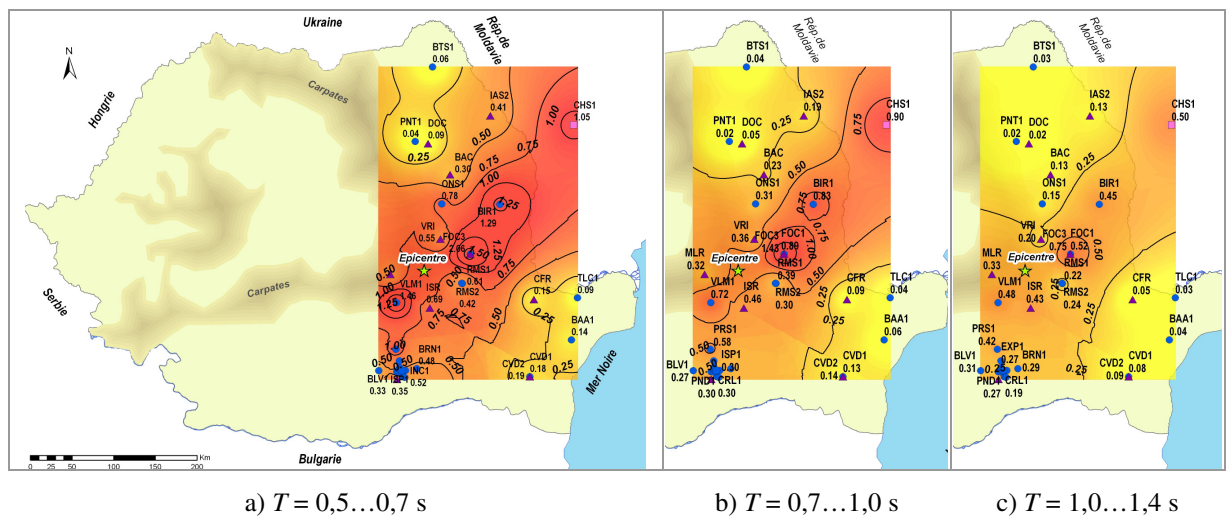

a) $T = 0,5…0,7$ s     b) $T = 0,7…1,0$ s     c) $T = 1,0…1,4$ s

**Figure 2.** *Cartes des valeurs spectrales de l'indicateur Park-Ang pour le séisme du 30 août 1986* ($C_y = 0,10$, $\mu_{mon} = 6$)

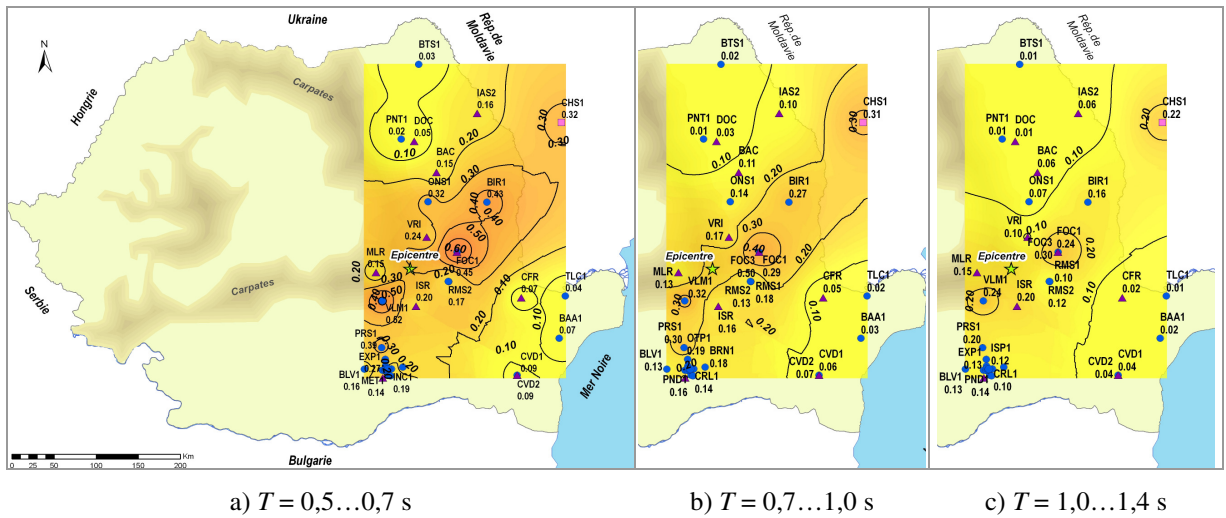

a) $T = 0,5…0,7$ s        b) $T = 0,7…1,0$ s        c) $T = 1,0…1,4$ s

**Figure 3.** *Cartes des valeurs spectrales de l'indicateur Park-Ang pour le séisme du 30 août 1986 ($C_y$ = 0,20, $\mu_{mon}$ = 6)*

Un exemple est celui de la fig. 2, où on observe des très grandes valeurs du *DM* aux stations VLM1, FOC3, BIR1 et CHS1. Pour le premier intervalle de période analysé, ces valeurs dépassent bien la limite de la ruine, en atteignant une valeur de 2,06 à Focsani (FOC3). Pour les autres intervalles de période, même avec la diminution avant-mentionnée de *DM* avec l'augmentation de la période, les valeurs de *DM* restent toujours très élevées à cette station. En termes strictes, ça signifie que, dans les villes concernées, les bâtiments de hauteur moyenne ($T$ =0.5…0.7 s) et de capacité de résistance modérée ($C_y$ = 0,10) auraient dû s'effondrer au séisme analysé. Heureusement, la situation réelle a été beaucoup moins grave, comme on montrera dans la dernière partie de l'article.

Les valeurs très élevées de *DM* qui résultent pour $C_y$ = 0,10 sont dues au caractéristiques du spectre de cet indicateur, mentionnées à la section 2.1. On pourrait tirer d'ici la conclusion que les estimations basées sur les valeurs spectrales de *DM* ne sont pas fiables pour des bâtiments rigides ou/et avec une capacité de résistance réduite ou modérée. En fait, même si les limitations inhérentes du calcul spectral jouent un rôle important dans cette déficience, la conclusion doit être nuancée en tenant aussi compte de l'influence de la sur-résistance sur les valeurs de $C_y$ (voir la formule [3]). Cet aspect est commenté dans la quatrième section de l'article.

## 3. Evaluation basée sur l'intensité sismique proposée par H. Sandi

### 3.1. *Définition de l'intensité sismique proposée*

L'intensité proposée par H. Sandi se détermine à partir des valeurs de réponse exprimée en accélérations absolues, $w_a(t, \varphi, 0,05)$, d'un pendule de fréquence propre (non-amortie) φ et de pourcentage d'amortissement critique de 5%, avec la formule (Sandi et al., 2011) :

$$i_d(\varphi) = \log_{7,5}\left(\int w_a^2(t, \varphi, 0,05)\, dt\right) + 6,45 \qquad [4]$$

Une description détaillé des raisons qui ont mené à la proposition de cette formulation de l'intensité et de ses avantages se trouve dans la référence bibliographique citée.

Pour évaluer le potentiel endommageant (la destructivité) sur des bandes de fréquences distinctes, on calcule les valeurs moyennes sur des bandes (φ', φ"), avec la formule

$$i_d*(\varphi',\varphi'') = \log_{7,5}\left\{\frac{1}{\ln(\varphi'',\varphi')}\int\left[\left(\int w_a^2(t,\varphi,0,05)\,dt\right)\frac{d\varphi}{\varphi}\right]\right\} + 6,45 \qquad [5]$$

On a choisi, pour cette étude, 3 bandes de fréquences, délimitées par les valeurs 0,707 Hz, 1 Hz, 1,414 Hz et 2 Hz. Elles sont sélectionnées parmi 12 intervalles divisant la bande de fréquences de 36 dB (0,25 Hz…16.0 Hz), considérée comme représentative. En utilisant pour les intensités moyennées une notation générique $I_{d12}$, suivie par le numéro de l'intervalle, les intensités qui correspondent respectivement aux 3 bandes choisies sont : $I_{d124}$ (0,707 Hz…1 Hz), $I_{d125}$ (1 Hz…1,414 Hz) et $I_{d126}$ (1,414 Hz…2 Hz). Si on exprime en périodes les 3 intervalles de fréquence, on observe qu'ils correspondent approximativement aux intervalles utilisés dans l'approche basée sur l'indicateur Park-Ang, à savoir (1.0…1.4 s), (0.7…1.0 s) et (0.5…0.7 s).

### 3.2. Cartes d'intensité sismique

En cartographiant les valeurs de l'intensité sismique Id12, correspondant aux 3 intervalles de période choisies, on obtient les cartes d'intensité sismique de la fig. 4.

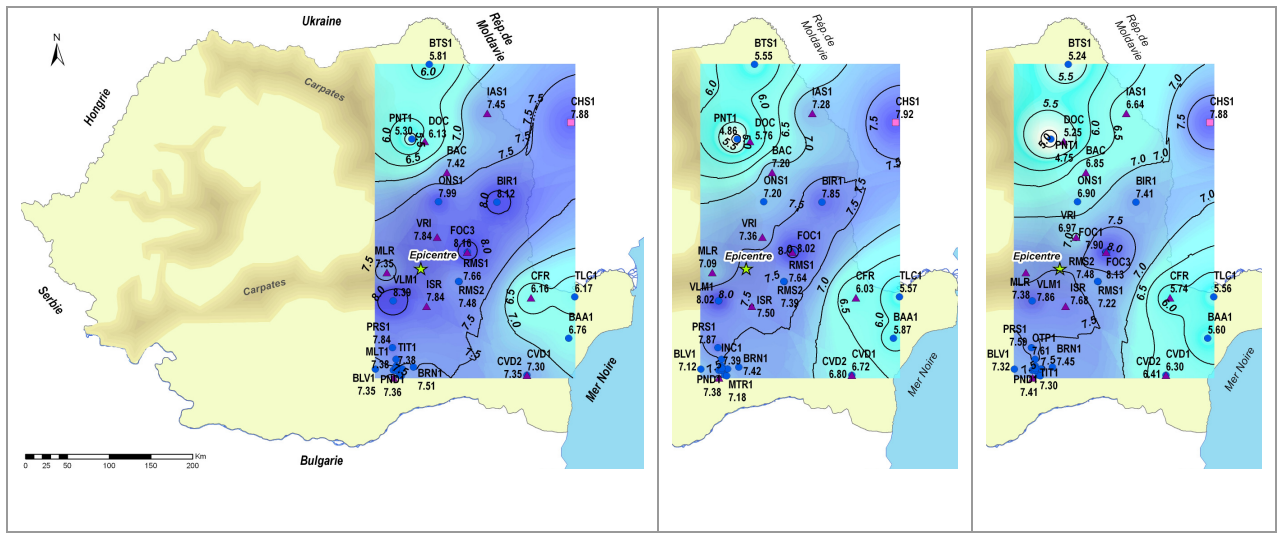

a) $T = 0,5…0,7$ s ($I_{d126}$)  b) $T = 0,7…1,0$ s ($I_{d125}$)  c) $T = 1,0…1,4$ s ($I_{d124}$)

**Figure 4.** *Cartes des valeurs de l'intensité sismique $I_{d12}$ pour le séisme du 30 août 1986*

Par l'analyse des cartes de la fig. 4 on remarque, pour les trois intervalles de période, une configuration des zones des valeurs maximales similaire à celle obtenue pour l'indicateur Park-Ang. Il convient, aussi, de remarquer la ressemblance de la forme des contours de la fig. 4 avec celle observée sur les cartes d'intensité macrosismique (Radu et al., 1986) réalisées après le séisme.

La plus grande valeur de l'intensité, $I_{d126}$ = 8,39, survient à Valenii de Munte (VLM1), pour $T$ = 0,5…0,7 s ($I_{d126}$). Pour le même intervalle, on trouve des valeurs supérieures à 8 également à Focsani (FOC3, $I_{d126}$ = 8,16) et à Barlad (BIR1, $I_{d126}$ = 8,12).

Comme tendance générale, l'intensité diminue pour les deux autres intervalles de période.

Si on analyse la variation des valeurs de l'intensité avec l'intervalle de période à chaque station, on peut faire les constatations suivantes :

- avec une seule exception, celle de la station INCERC Bucarest (INC1), les plus grandes valeurs de l'intensité surviennent pour l'intervalle $T$ = 0,5…0,7, donc pour $I_{d126}$ ;
- à la station INCERC Bucarest, l'intensité maximale survient pour l'intervalle $T$ = 1,0…1,4 s, $I_{d126}$ = 7,64; pourtant, les différences entre les valeurs enregistrées pour les trois intervalles sont seulement d'environ 3% ;
- les valeurs minimales apparaissent aux stations situées au nord-ouest et au sud-est de la ligne VLM1-CHS1 ;
- la variation de l'intensité avec la période n'est pas monotone, car elle porte l'empreinte du contenu de fréquences du mouvement sismique ; donc, l'intervalle de périodes dans lequel on enregistre l'intensité maximale diffère d'une station à l'autre ; pourtant, dû au fait que la majorité des mouvements sismiques analysés sont à bande large de fréquences, l'allure générale des contours varie assez peu avec la période.

## 4. Comparaisons et commentaires

L'analyse de cette section est focalisée sur les bâtiments en béton armé, dimensionnes conformément aux normes sismiques roumaines antérieures au séisme de 30 août 1986, à savoir P13-63, P13-70, P100-78 et P100-81. Tenant compte des bâtiments existants à l'époque en Roumanie, les types de structures sont présentées dans le tableau suivant. Puisque le troisième intervalle de périodes correspond aux bâtiments de hauteur très élevée, atypiques pour le parc de logements courant, on analyse seulement les deux premiers intervalles. Pour la classification des structures, on a utilisé les types spécifiés par l'échelle d'intensité EMS-98 (Grünthal, 1998). Les périodes sont calculées avec les formules simplifiées fournies par les normes roumaines considérées.

En analysant les résultats des sections précédentes par le prisme des valeurs du tableau 1, on peut faire la discussion suivante.

En se rapportant d'abord á l'approche basée sur l'indicateur Park-Ang, on considère, en guise d'exemple, un coefficient moyen de sur-résistance $R_{OVS}$= 2 (Craifaleanu, 2010). On observe que, dans ce cas, toutes les valeurs du $C_s$ prescrites pour Bucarest par les normes P13-63 et P13-70, conduisent, selon la formule [3], à des valeurs $C_y$ inferieures à 0,10. Seulement les valeurs prescrites par les normes P100-78 et P100-81 conduisent aux valeurs $C_y$ supérieures à 0,10. Quant aux valeurs qui résultent pour la ville de Focsani, pour le même $R_{OVS}$ considéré on atteint et même on dépasse la valeur $C_y$ = 0,20, à l'exception de la valeur de 0,060 correspondant aux ossatures d'hauteur moyenne dimensionnées selon la norme P13-70. Ces résultats montrent une situation très défavorable à certaines stations de Bucarest, où les valeurs $DM$ calculées pour ces coefficients $C_y$ et pour une ductilité $\mu_{mon}$ = 6 sont supérieures à la limite des endommagements réparables ($DM$ = 0,4). En réalité, selon les rapports de l'époque, les endommagements enregistrés à Bucarest aux structures de résistance des bâtiments ont été mineurs et isolés.

Une estimation correcte de l'endommagement nécessite une évaluation plus précise du coefficient de sur-résistance $R_{OVS}$ dans la formule [3]. Par exemple, dans le cas des bâtiments rigides, les valeurs de la sur-résistance peuvent être supérieures à celles propres aux bâtiments flexibles, due aux mesures constructives imposées par les normes. Pour un $C_s$ donné, cela conduira aux valeurs de $C_y$ plus grandes dans le cas des bâtiments rigides, donc à une diminution des valeurs spectrales de $DM$ obtenues dans la zone des périodes courtes. Par conséquent, on obtiendra ainsi une estimation de l'endommagement plus proche de celle constatée en réalité.

**Tableau 1.** *Types de structures analysés*

| $T$ (s) | Type de structure | Niveau de conception parasismique | Hauteur (niveaux) | Classe de vulnérabilité EMS-98 | Norme | Coefficient sismique $C_s$ (selon les normes de l'époque) | |
|---|---|---|---|---|---|---|---|
| | | | | | | Bucarest | Focsani |
| 0,5…0,7 | Ossatures en b.a. | moyen | 4…7 | B, C, **D**, E | P13-63 | 0,036 | 0,144 |
| | | | | | P13-70 | 0,032 | 0,085 |
| | | bon | | C, D, **E**, F | P100-78, P100-81 | 0,064 | 0,102 |
| | Murs en b.a. | moyen | 9…13 | C, **D**, E | P13-63 | 0,028 | 0,113 |
| | | | | | P13-70 | 0,036 | 0,096 |
| | | bon | | D, **E**, F | P100-78, P100-81 | 0,075 | 0,120 |
| 0,7…1,0 | Ossatures en b.a. | moyen | 7…10 | B, C, **D**, E | P13-63 | 0,025 | 0,102 |
| | | | | | P13-70 | 0,023 | 0,060 |
| | | bon | | C, D, **E**, F | P100-78, P100-81 | 0,064 | 0,102 |

En ce qui concerne les évaluations basées sur l'intensité instrumentale de la section 3 de l'article, les valeurs de $I_{dI2}$ obtenues pour Bucarest se chiffrent entre 7,09 et 7,80. En considérant une intensité de 8, cela correspond, sur l'échelle EMS-98, à des dégâts de degré 3 aux bâtiments de la classe de vulnérabilité B, et à des dégâts de degré 2 aux bâtiments des classe de vulnérabilité C et D. Selon les rapports de l'époque, les endommagements enregistrés dans la capitale aux structures de résistance des bâtiments correspondent, dans le cas des types de structures du tableau 1, à des dégâts de degré 2 (Grünthal, 1998). Tenant compte que la classe de vulnérabilité B pour les ossatures du tableau 1 est qualifiée comme un cas exceptionnel, très peu probable, on peut conclure que, pour Bucarest, l'évaluation basée sur l'intensité instrumentale est absolument satisfaisante. En ce qui concerne la ville de Focsani, les intensités calculées se situent entre 7,90 et 8,16, ce qui conduit à une évaluation des dégâts légèrement supérieure à celle réalisée pour Bucarest. Par comparaison avec les rapports de l'époque, l'évaluation est satisfaisante pour les bâtiments en béton armé considérés dans cette étude. Il convient de remarquer qu'à Focsani on a aussi enregistré, au séisme du 30 août 1986, des dégâts importants (de degré 3…4) à deux bâtiments de patrimoine, avec des structures en maçonnerie, encadrées dans la classe de vulnérabilité A, **B** (la lettre grasse signifie la valeur la plus probable, selon l'échelle EMS-98). Cela correspond aux résultats de l'évaluation basée sur l'intensité instrumentale.

## 5. Conclusions

L'évaluation du potentiel endommageant sur la base des valeurs spectrales de l'indicateur Park-Ang a conduit, spécialement pour les forts tremblements de terre, aux estimations trop sévères dans le cas des structures à périodes courtes et/ou à capacité de résistance réduite ou modérée. Une évaluation plus détaillée de la sur-résistance des structures peut améliorer la précision des estimations.

L'évaluation du potentiel endommageant sur la base de l'intensité sismique instrumentale proposée par H. Sandi a démontré une très bonne correspondance avec les dégâts enregistrés effectivement aux bâtiments, due à sa calibration par rapport à l'échelle EMS-98.

## 6. Bibliographie